\documentclass[a4paper,11pt]{article}
\usepackage{jinstpub} 

\usepackage{graphicx}
\graphicspath{{figures/}}
\usepackage{tikz}
\usepackage{float}
\usepackage{pdfpages}
\usepackage{caption}
\usepackage{subcaption}
\usepackage{makecell}
\usepackage{booktabs}
\usepackage{multirow}
\usepackage{textcomp}
\usepackage{siunitx}

\title{Development, Characterization, and Testing of a Bias Supply for SiPMs in the CMVD Experiment}

\author[1]{Prajjalak Chattopadhyay,
\note{Corresponding author.}}
\author{Mandar N. Saraf,}
\author{Gobinda Majumder,}
\author{Satyanarayana Bheesette,}
\author{and Ravindra R. Shinde}
\affiliation{Tata Institute of Fundamental Research,\\
Mumbai, India}

\emailAdd{prajjalak.chattopadhyay@tifr.res.in}

\abstract{To assess the viability of a shallow-depth neutrino detector, a Cosmic Muon Veto Detector (CMVD) is being constructed on top of the stack of Resistive Plate Chamber (RPC) detectors at TIFR, Mumbai. The CMVD employs extruded plastic scintillators for muon detection, with wavelength-shifting fibers coupled to silicon photomultipliers (SiPMs) for signal readout. A highly stable, low-noise power source is essential for biasing the SiPMs, as the precision, accuracy, and stability of the supply directly impact the consistency of their gain. To address this, we designed a biasing power supply capable of delivering \SIrange{50}{58}{\volt} in \SI{50}{\milli\volt} steps, with a maximum short-circuit current output of \SI{1}{\milli\ampere}. The system incorporates digital voltage control, stabilization, and current monitoring, making it compatible with external controllers (such as microcontrollers). This added flexibility and modularity allow for additional functionalities, including temperature compensation. Designed to supply multiple SiPMs with close to breakdown voltages in parallel, the circuit seamlessly integrates with the front-end electronics of the detector system.}


\arxivnumber{2505.04510}

\begin{document}
\maketitle
\flushbottom

\section{Introduction}
The silicon photomultiplier (SiPM) is a multi-pixel photo detector where each pixel is in Geiger mode and as a whole works in avalanche mode \cite{NIMA926}. It has a high gain ($10^5$ - $10^6$), good photon detection efficiency, compact size (a few millimeters), low cost, and is insensitive to magnetic field. However, it has a high noise rate ($\sim$\SI{100}{\kilo\hertz}) and a higher nonlinearity compared to photomultiplier tubes (PMTs).

A tall mountain providing more than \SI{1.3}{\kilo\meter} of rock overburden -- equivalent to approximately 3500 meter water equivalent (mwe) in all directions -- can suppress atmospheric muon flux by nearly six orders of magnitude. This level of background reduction is essential for rare event searches, such as cosmic neutrino detection, where neutrinos are identified indirectly via the muons produced in their interactions with the target medium. Given that only a few such neutrino interactions may occur per day, minimizing the background from atmospheric muons is crucial to achieve a favorable signal-to-background ratio. Although such deep underground sites are ideal, a shallower facility with a high-efficiency active muon veto system could make many more sites available. For example, with a veto efficiency of 99.99\%, a facility at a depth of around \SI{100}{\meter} could be equivalent, and has already been conceptualized by Shah et al. \cite{Shah_2024}. To explore this possibility, a Cosmic Muon Veto Detector (CMVD) \cite{saraf_CMVD_2024} is being constructed on top of the existing 12 layer Resistive Plate Chamber (RPC) \cite{RPC_Santonico} detector stack at the Tata Institute of Fundamental Research, Mumbai, India. RPCs are gaseous particle detectors composed of parallel resistive electrodes separated by a narrow gas-filled gap, where ionization by charged particles initiates avalanche multiplication. Known for their excellent time resolution, RPCs are widely used in triggering and timing applications. In the CMVD setup, the RPC stack will serve both as an external trigger source and as an external muon tracking system, and will be used to determine the efficiency of the CMVD.

\begin{figure}[htbp]
    \centering
    \begin{subfigure}[b]{0.45\textwidth}
        \centering
        \begin{tikzpicture}
            \node[anchor=south west, inner sep=0] (image) at (0,0)
            {\includegraphics[width=\textwidth]{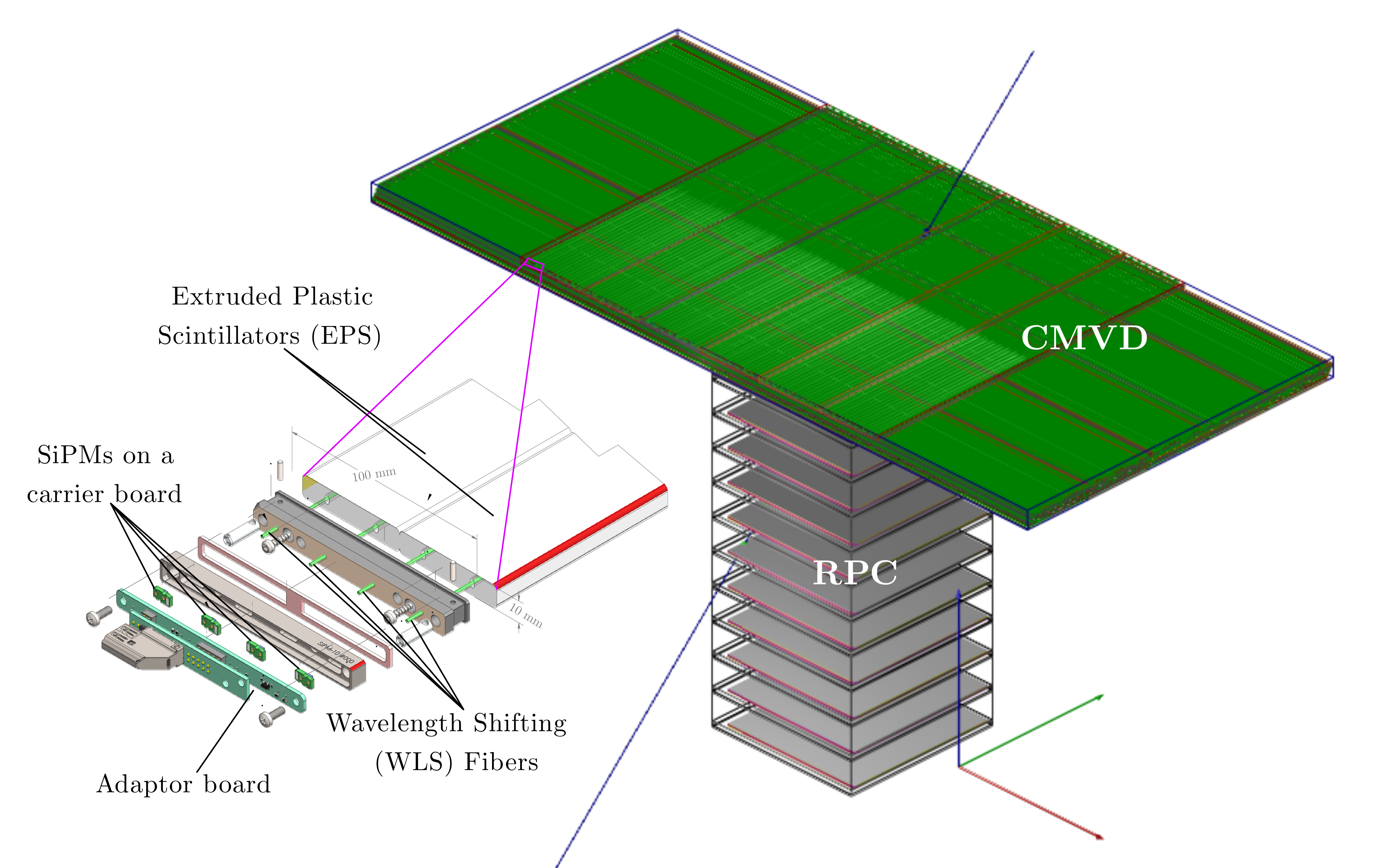}};
            \begin{scope}[x={(image.south east)}, y={(image.north west)}]
                \node[red!60!black, font=\tiny] at (0.81, 0.02) {x};
                \node[green!60!black, font=\tiny] at (0.81, 0.21) {y};
                \node[blue!60!black, font=\tiny] at (0.69, 0.35) {z};
                \draw[<->, red!60!black, thick] (0.505, 0.13) -- (0.605, 0.05);
                \node[red!60!black, font=\tiny] at (0.54, 0.06) {\SI{1}{\meter}};
                \draw[<->, green!60!black, thick] (0.61, 0.05) -- (0.72, 0.14);
                \node[green!60!black, font=\tiny] at (0.68, 0.06) {\SI{1}{\meter}};
                \draw[<->, magenta!80!black, thick] (0.6, 0.9) -- (0.86, 0.7);
                \node[magenta!80!black, font=\tiny] at (0.775, 0.82) {\SI{2.4}{\meter}};
                \draw[<->, red!60!black, thick] (0.575, 1.03) -- (1.05, 0.665);
                \node[red!60!black, font=\tiny] at (0.84, 0.88) {\SI{4.5}{\meter}};
                \draw[<->, green!60!black, thick] (0.75, 0.365) -- (0.96, 0.535);
                \node[green!60!black, font=\tiny] at (0.87, 0.41) {\SI{2}{\meter}};
                \draw[<->, blue!60!black, thick] (0.97, 0.17) -- (0.97, 0.55);
                \node[blue!60!black, font=\tiny] at (1.005, 0.3) {\SI{2}{\meter}};
            \end{scope}
        \end{tikzpicture}
        \caption{}
    \end{subfigure}
    \hspace{.05\textwidth}
    \begin{subfigure}[b]{0.46\textwidth}
        \centering
        \begin{tikzpicture}
            \node[anchor=south west, inner sep=0] (image) at (0,0)
            {\includegraphics[width=\textwidth]{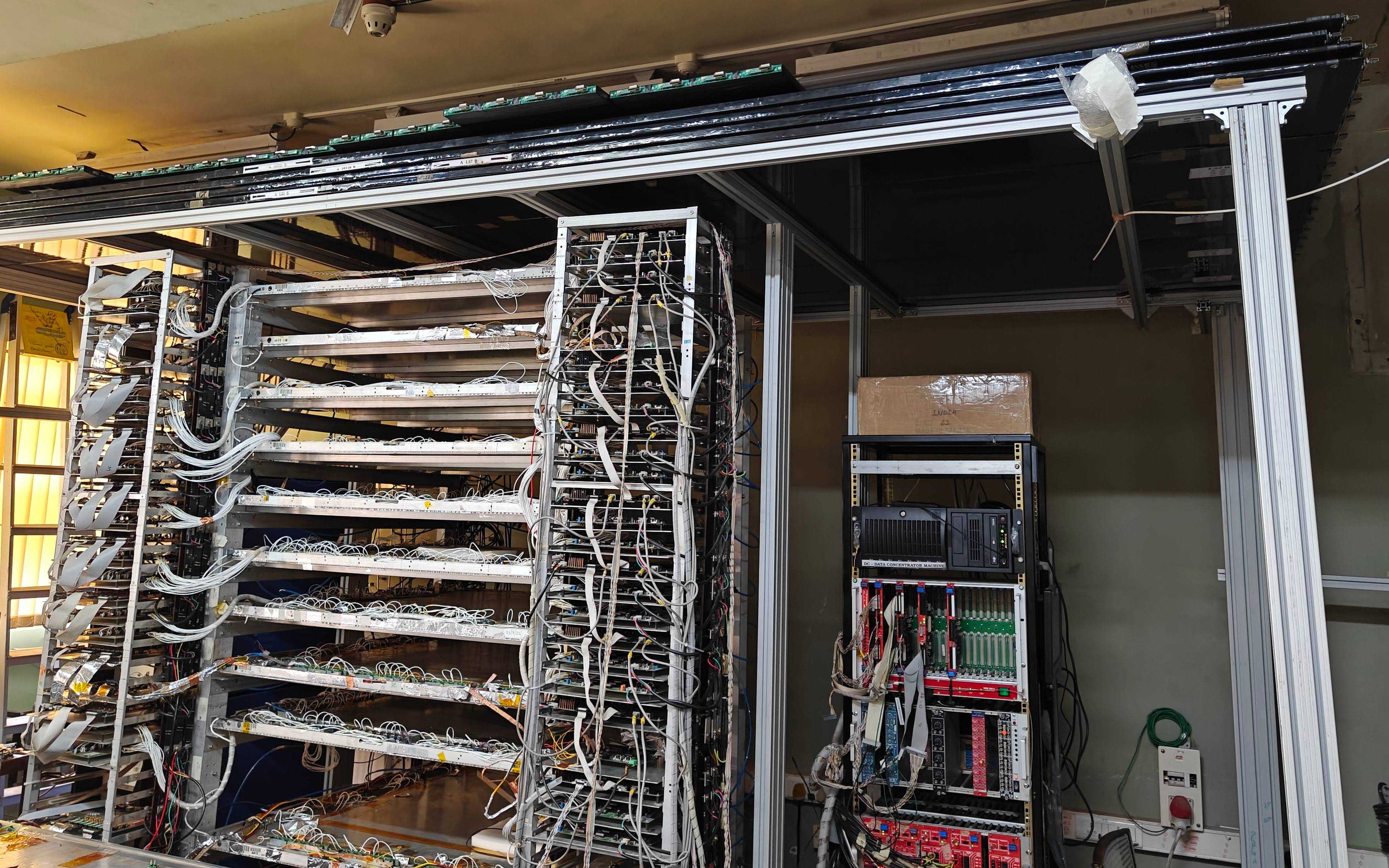}};
            \begin{scope}[x={(image.south east)}, y={(image.north west)}]
                \node[fill=white, text=black, font=\footnotesize] at (0.3, 0.5) {RPC};
                \node[white, font=\footnotesize] at (0.6, 0.9) {CMVD};
                \draw[<-, white, thick] (0.42, 0.9) -- (0.38, 0.93);
                \node[white, font=\footnotesize] at (0.36, 0.95) {SiPM assembly};
            \end{scope}
        \end{tikzpicture}
        \caption{}
    \end{subfigure}
    \caption{The Cosmic Muon Veto Detector on top of the RPC stack. (a) active components with a muon trajectory (blue) in Geant4 \cite{Geant4_1, Geant4_2, Geant4_3}. The green shadowed region represents the active detection volume of the scintillators. The scintillators are placed in  layers, where the bottom 4 layers are along the x-direction, and the top layer along y-direction. The placement of the SiPMs has been shown as well. (b) view of the real detector.}
    \label{fig:CMVD_photo}
\end{figure}

The CMVD is shown in figure \ref{fig:CMVD_photo}, which consists of extruded plastic scintillator tiles (EPS) with wavelength shifting (WLS) fibers running inside the EPS through holes that were made during the extrusion process \cite{saraf_EPS_2023}. It has five \SI{400}{\milli\meter} wide and \SI{4.5}{\meter} long scintillator tiles along the x direction in 4 layers, and six \SI{400}{\milli\meter} wide and \SI{2}{\meter} long scintillator tiles along the y direction in a single layer. There are 16 fibers in each tile and the fibers are terminated at the ends of the tile, where a SiPM (Hamamatsu S13360-2050VE) is connected to each fiber. Four such SiPMs are mounted on a carrier board, which distributes power to them and routes their output signals to an HDMI port. The 4 differential pairs of the HDMI cable carry 4 signals, one pair for each SiPM. The HDMI cable also carries the power for all SiPMs. Each tile has four such boards on each side, resulting in a total of 32 SiPMs.

To power the SiPMs, a very precise, accurate, low noise, and stable power supply is required. This article focuses on the design, testing, and characterization of such a power supply.

\section{Operating considerations of SiPMs\label{sec:theory}}
When a photon enters an SiPM, it generates an electron--hole (e--h) pair in the depletion layer of a reverse-biased p--n junction. In Geiger mode, because of a strong electric field, the initial e--h pair triggers an avalanche, producing a measurable charge. The avalanche is quenched by an integrated quenching resistor. The total charge generated per avalanche is \cite{corsi_modelling_2007}:

\begin{equation}
    Q = C \cdot V_{OV}(T)
    \label{eqn:charge}
\end{equation}

Here, $C$ represents the effective capacitance of the SiPM pixel, incorporating both the intrinsic junction capacitance as well as additional parasitic contributions. The term $V_{OV}(T)$ represents the overvoltage at temperature $T$, and is defined as the difference between the applied bias voltage $V_{BIAS}$ and the breakdown voltage $V_{BD}(T)$, $V_{OV}(T) = V_{BIAS} - V_{BD}(T)$.

The gain $G$, defined as the ratio of output charge to elementary charge $q_e$, in the range of $V_{OV}(T)$ from \SIrange{0.2}{6}{\volt}, is \cite{marrocchesi_active_2009}:

\begin{equation}
    G(T) = \frac{C \cdot V_{OV}(T)}{q_e} = G_{0} \cdot V_{OV}(T)
    \label{eqn:gain}
\end{equation}

Where $G_{0}$ is the gain per volt of overvoltage. The overvoltage not only determines the amount of charge (and in turn the gain), but also influences the photon detection efficiency (PDE), dark count rate (DCR), and crosstalk probability of the SiPM, as shown by Jangra et al. \cite{Mamta_1}. Therefore, precise control and stability of $V_{OV}$ are crucial for consistent detector performance.

A key consideration in SiPM operation is the strong temperature dependence of the breakdown voltage. The breakdown voltage $V_{BD}$ increases linearly with temperature, typically modeled as:

\begin{equation}
    V_{BD}(T) = a_{0} (T - T_{0}) + V_{0}
\end{equation}

where $a_{0}$ is the temperature coefficient of the breakdown voltage and $V_{0}$ is the breakdown voltage at some temperature $T_{0}$. For the SiPM used in this work (Hamamatsu S13360-2050VE), the value of $a_{0}$ is specified as \SI{54}{\milli\volt/\degreeCelsius} at $T_{0}$\,=\,\SI{25}{\degreeCelsius} and $V_{OV}$\,=\,\SI{3}{\volt} \cite{hamamatsu_photonics_mppc_2024}.

In practical terms, this temperature dependence makes the gain highly sensitive to fluctuations in either the ambient temperature or the applied bias voltage. For example, a \SI{1}{\milli\volt} variation in the bias voltage at $V_{OV}$=\SI{1}{\volt} can lead to a gain change of 0. 1\%. Therefore, it becomes essential to use a well-regulated, low-noise, and thermally stable power supply to maintain constant overvoltage and thereby ensure reliable operation of the SiPM.

In addition to hardware stabilization, many systems employ active compensation methods, such as closed-loop feedback control using temperature sensors, which adjust the bias voltage in real time to maintain constant overvoltage and thus the gain as temperature drifts; Kumar et al., for example \cite{Hagar_bias_supply_2023}. While such solutions are effective and are necessary for many applications, in the scope of this work, temperature compensation is not a necessity because the detectors will be in a temperature-controlled environment. This forms the primary motivation for the design and development of the custom SiPM bias supply system discussed in this work.

\section{Requirements of the power supply\label{sec:requirements}}
In this design, one of the key aspects was to group 32 SiPMs with closely matched breakdown voltages. This is essential because the front-end readout electronics do not have a HV trimming facility. Therefore, it is necessary to ensure that all 32 SiPMs in a group would have a consistent gain under a common biasing voltage.

To accommodate this requirement, the system is needed to provide a programmable bias voltage ranging from \SIrange[range-units = single]{50}{56}{\volt} direct current (DC). This range was chosen on the basis of the measured breakdown voltages of the SiPMs that were procured from the manufacturer. Moreover, fine control over this bias voltage was critical for tuning and matching the operating points. Hence, the design aimed for a minimum control step size of \SI{50}{\milli\volt} and a power supply noise of \SI{10}{\milli\volt} or better, allowing precise adjustment and regulation of the bias voltage.

In terms of current capacity, the power supply was required to support up to \SI{25}{\micro\ampere}. This value is more than ten times the expected steady-state current drawn by the group of SiPMs, offering ample headroom for start-up transients and any potential leakage current variations. At the same time, the system needed to monitor the current with high resolution to detect subtle changes that may indicate faults or issues such as increased noise or thermal effects. For this purpose, a current readout with a least count of \SI{50}{\nano\ampere} or better was recommended.

Several additional features were considered essential for safe and reliable operation. These included over-current protection to prevent device damage in case of accidental faults, and a soft-start mechanism to avoid sudden voltage surges during power-up. Furthermore, the ability to digitally control the output voltage and read the output current was necessary for integration into the broader data acquisition and control infrastructure. These digital interfaces would enable remote configuration, monitoring, and logging -- key requirements in complex detector systems.

\section{Electronics design\label{sec:design}}
\begin{figure}[htbp]
    \centering
    \begin{subfigure}[b]{0.45\textwidth}
        \centering
        \includegraphics[width=\textwidth]{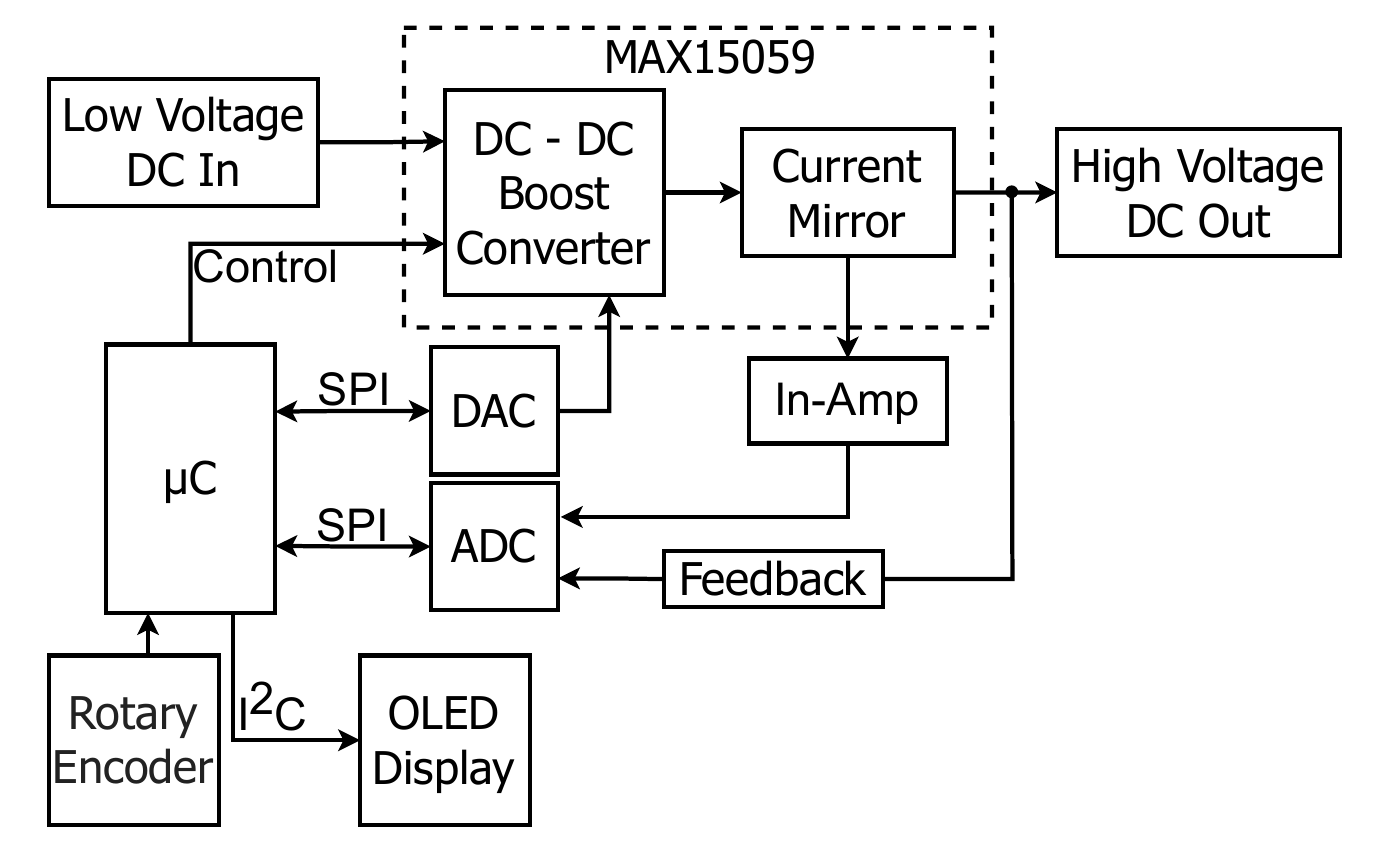}
        \caption{}
    \end{subfigure}
    \hspace{.02\textwidth}
    \begin{subfigure}[b]{0.5\textwidth}
        \centering
        \includegraphics[width=\textwidth]{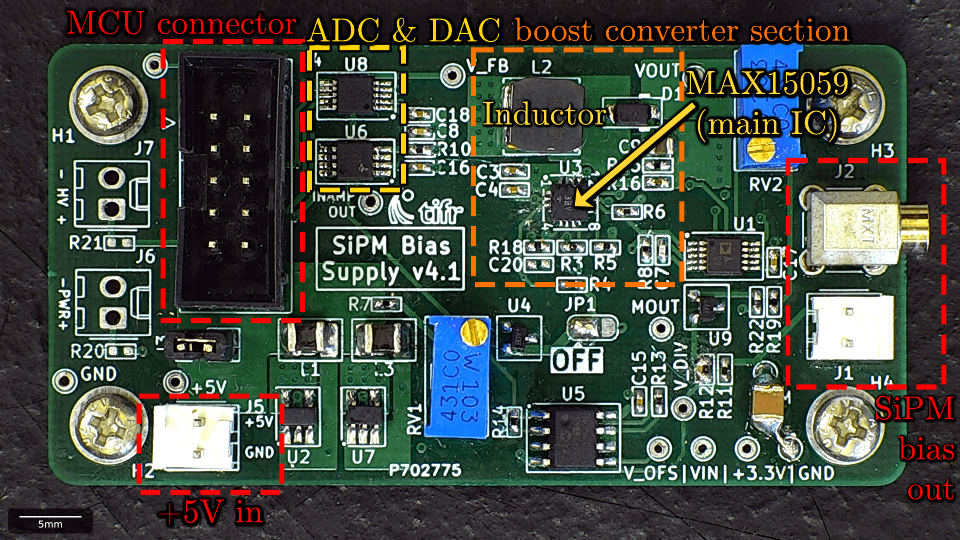}
        \caption{}
    \end{subfigure}
    \caption{(a) Block diagram of the circuit. (b) Fully populated PCB of the SiPM bias supply.}
    \label{fig:block-diag_pcb}
\end{figure}

The circuit design in this work is derived from the architecture proposed by Gil et al. \cite{gil_programmable_2011}, with significant modifications. While their implementation is based on a commercially available power supply module, our approach involves a complete in-house development of both the DC--DC boost converter and the associated control electronics. The general circuit architecture is shown in figure \ref{fig:block-diag_pcb}(a), and a photograph of the fully assembled PCB is shown in figure \ref{fig:block-diag_pcb}(b).

Here, the boost converter is built around the MAX15059 IC. This converter generates a programmable higher voltage output from a lower DC input, providing the necessary bias voltage for the intended application. To enable precise voltage regulation, the system employs a closed-loop feedback mechanism. The output voltage and current are continuously monitored by a 12-bit Successive Approximation Register (SAR) Analog-to-Digital Converter (ADC). The voltage is first attenuated using a resistive voltage divider and then passed through an op-amp-based analog conditioning stage that scales the signal to match the ADC's input range. This whole section is shown as the \emph{Feedback} block in figure \ref{fig:block-diag_pcb}(a). Digitized values are sent to an Arduino Pro Mini microcontroller unit (MCU or \textmu C), which executes a Proportional--Integral--Derivative (PID) control algorithm \cite{PID_book}. The algorithm calculates a correction signal, defined by the control law:

\begin{equation}
    u(t) = K_p e(t) + K_i \int_0^t e(\tau) d\tau + K_d \frac{de(t)}{dt}
    \label{eq:PID}
\end{equation}

Here, $e(t)$ is the error term, defined as the difference between the set point and the measured input at time $t$: $e(t) = y_{set} - y_{meas}(t)$. The terms $K_p$, $K_i$, and $K_d$ are the PID control parameters, called proportional, integral, and differential coefficients, respectively. The integral term accumulates the error from the start of the device to the current time $t$, as shown in the limits of the integral. $u(t)$ is the output of the PID controller, which is sent to a 12-bit Digital-to-Analog Converter (DAC). This produces an analog voltage at the output of the DAC that acts as a control voltage for the power supply. This voltage drives the internal feedback network of the MAX15059 through a resistor divider, thus modulating the output voltage accordingly. These resistors can be changed to obtain different output voltage ranges, from $V_{CC}$\,+\,\SI{5}{\volt} to \SI{76}{\volt}, according to the datasheet of the IC. Here, $V_{CC}$ is the input voltage for the IC.

Additionally, the MAX15059 features an internal current mirror that outputs a scaled replica of the output current. This signal is routed through an instrumentation amplifier (in-amp) for amplification and then digitized by the second channel of the ADC. Since the current is very low (in the order of nA), a general-purpose op-amp will skew the current readout significantly and hence cannot be used in this case. Due to its significantly lower input offset voltage and current, as well as high input impedance, an in-amp is used in this case, which will not skew the readout significantly. The current readout is primarily intended for monitoring and safety diagnostics, but can also be used for further control strategies if required.

User interaction is facilitated through a rotary encoder and an OLED display. Rotating the encoder adjusts the target output voltage, while the built-in push button is used to set the selected value and turn the output on. The OLED screen provides real-time visual feedback of the set voltage, measured output voltage, and output current, allowing the user to intuitively monitor the behavior of the system. The design also supports the possibility of replacing the manual input with digital input using the communication channel for a fully automated operation. Moreover, the PID controller itself is not limited to the Arduino platform; Any device capable of implementing a PID algorithm, for example, an MCU or FPGA, could be integrated into this framework.

The hardware is implemented on a custom-designed 4-layer printed circuit board (PCB) as shown in figure \ref{fig:block-diag_pcb}(b), designed using KiCad EDA software. The board includes an internal ground plane for noise reduction and two distinct power planes to isolate signal and power sections. Signal traces are routed on the top and bottom layers, with most components placed on the top layer and a few passive elements on the bottom. A dedicated connector links the main board to a detachable MCU sub-board, enabling modularity and system-level flexibility. This modular design allows the system to be easily scaled for larger setups such as the CMVD, where multiple of such power supplies will operate simultaneously. In such configurations, a single controller with multiple SPI buses can manage several units, and additional features like temperature compensation or remote reconfiguration can be incorporated with minimal modification.

\section{Calibration}
\subsection{PID tuning}
\begin{figure}[htbp]
    \centering
    \includegraphics[width=\textwidth]{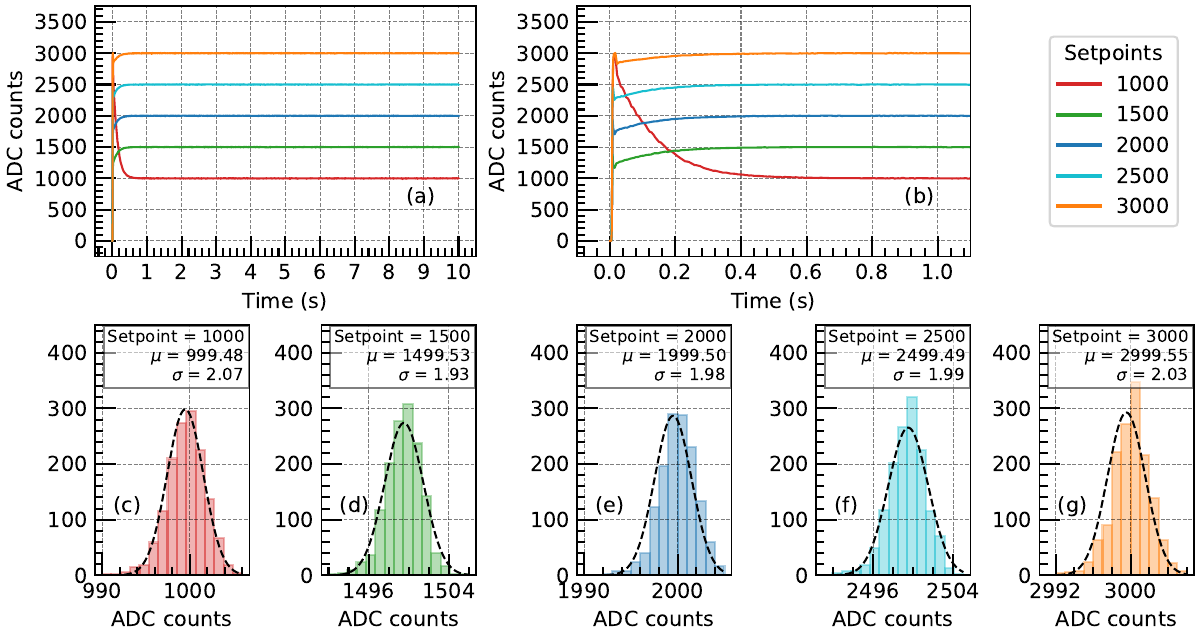}
    \caption{(a) Step response after PID tuning. (b) Zoomed-in view of the time axis that corresponds to the initial transient response. (c)-(g) Distribution of ADC readout values after stabilization with a Gaussian fit for different setpoints.}
    \label{fig:PIDcal}
\end{figure}

The first step of the device calibration process was to configure the PID controller, where we tuned the PID parameters described in equation \ref{eq:PID}. For that we used a combination of the Ziegler-Nichols tuning method \cite{microstar_laboratories_inc_ziegler-nichols_nodate} and the MATLAB PID Tuner app \cite{PID_tuner_matlab}. During the tuning process, a step function was sent as an input and response of the output was observed more than 50 times by changing the PID parameters, until a good set of parameters was found. Once the parameters were determined, more than 10 of these data were recorded for plotting, sometimes with tiny variation in the control parameters. Some of the responses of the PID controller after tuning is shown in figure \ref{fig:PIDcal}(a), for different input step function heights (setpoints). The transient response at the beginning is shown in figure \ref{fig:PIDcal}(b). Figures \ref{fig:PIDcal}(c)-(g) show the distribution of ADC counts once the initial transient period is over and the system enters the stable state, i.e., the region from \SI{1.5}{\second} to \SI{10}{\second} in the plot \ref{fig:PIDcal}(a). The response shows a standard deviation of roughly 2 ADC counts, as determined from figure \ref{fig:PIDcal}(c)-(g). This implies roughly \textpm\SI{6}{\milli\volt} peak-to-peak ripple noise, as determined from the voltage calibration in section \ref{sec:Vcal}.

\subsection{Voltage calibration\label{sec:Vcal}}
Since the output voltage is sampled by an ADC after some analog signal conditioning, a calibration must be performed to map the output voltage values to the ADC counts. For that purpose, a Keithley 2450 SMU \cite{keithley_2450} has been used as a reference to measure the output voltage. The calibration process has been automated by connecting the SMU and the MCU of the bias supply to a PC and using a Python script with the PyVISA \cite{pyVISA} and PySerial \cite{pyserial} libraries to read the SMU and the MCU, respectively.

\begin{figure}[htbp]
    \centering
    \includegraphics[width=\textwidth]{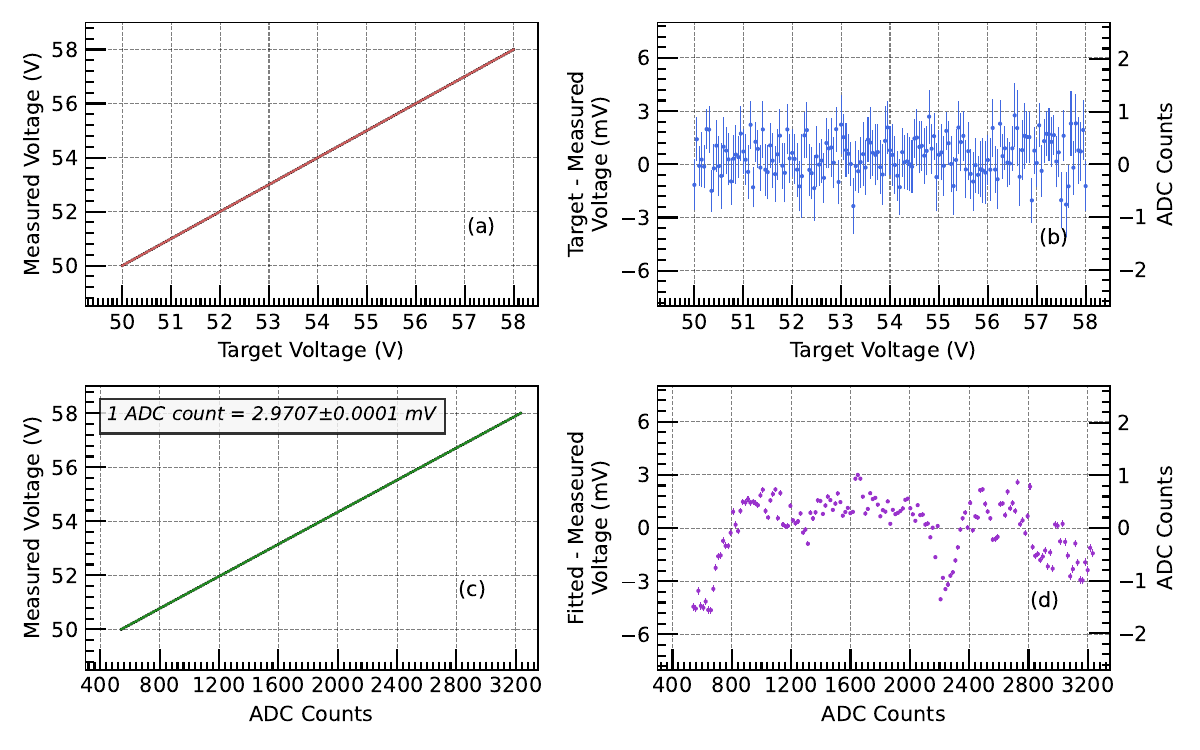}
    \caption{(a) Measured output voltages vs set output voltages with a linear fit. (b) Difference between measured output voltages and set voltages, where error bars represent the standard deviation of the measurements. (c) Measured output voltage vs ADC counts with a linear fit, which we call the voltage calibration. (d) Residuals of the linear fit.}
    \label{fig:Vcal}
\end{figure}

The set voltages vs measured voltages are shown in figure \ref{fig:Vcal}(a). The differences in set and measured voltages are shown in figure \ref{fig:Vcal}(b), where the error bars represent the standard deviation in measured voltages. It is clear that the deviations are within the \textpm 1 ADC counts, where the 1 ADC count is the smallest quantization step. The voltage calibration data are shown in figure \ref{fig:Vcal}(c). The residuals plot (difference between the fitted value and the actual data point) is shown in figure \ref{fig:Vcal}(d), where the error bars represent the error in fitting.

The calibration was done in \SI{50}{\milli\volt} steps because this has been kept as the step size of the voltage setting. The calibration data are the value of the ADC codes corresponding to each voltage value, which are uploaded to the internal EEPROM of the MCU as a look-up table. This method, over a linear approximation method, ensures that any systemic nonlinearity will be taken care in the calibration data, and single-point nonlinearities are compensated properly.

\subsection{Current readout calibration}
Finally, a current readout calibration was performed to translate the ADC counts into actual current values. For this, again, the Keithley 2450 SMU was used as a programmable constant current dummy load. This can be achieved by using the current limit function of the SMU in the source V measure I configuration but sourcing zero volts.

\begin{figure}[htbp]
    \centering
    \includegraphics[width=\textwidth]{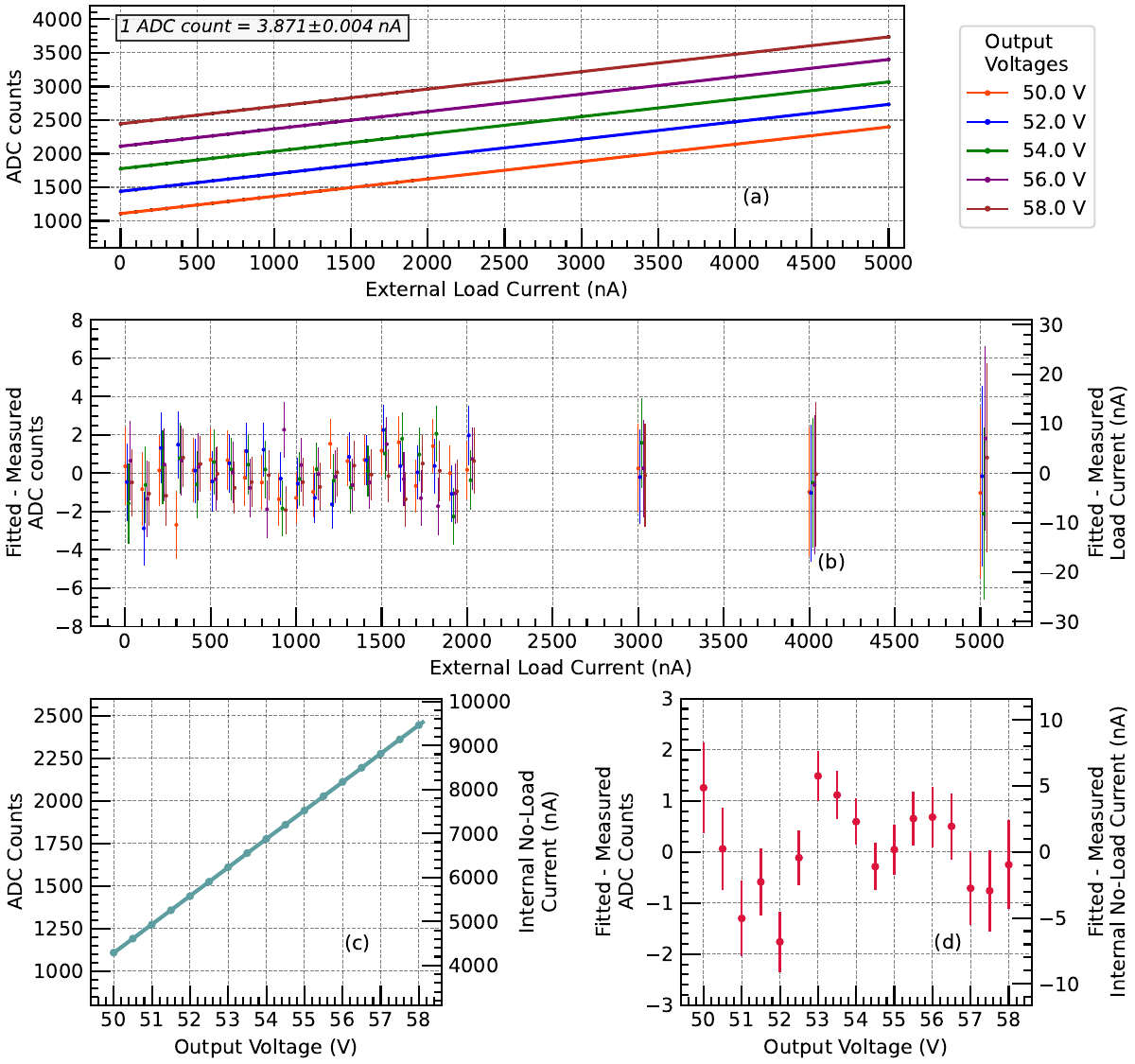}
    \caption{(a) Current calibration with a linear fits for different output voltages. (b) Residuals of the linear fittings of (a). (c) No-load current as a function of the output voltage, with a linear fit. (d) Residuals of the linear fit in (c).}
    \label{fig:Ical}
\end{figure}

The results of the current calibration are shown in figure \ref{fig:Ical}(a), where different load currents were applied for different voltages, and the points were fitted with straight lines. The residuals of the fits are shown in figure \ref{fig:Ical}(b), where the deviation is mostly within \textpm \SI{10}{\nano\ampere}, which is also within the acceptable limits. Since the output line has the resistive voltage divider as a part of the voltage feedback loop, the no-load current is dependent on the output voltage, following Ohm's law. Therefore, the current readout also depends on the output voltage. The no load current as a function of the output voltage along with a linear fit is shown in figure \ref{fig:Ical}(c). The slope and intercept of this line will allow us to estimate the amount of offset that needs to be subtracted for a certain output voltage. Figure \ref{fig:Ical}(d) shows the residuals of the fit of figure \ref{fig:Ical}(c). From the least count (shown in \ref{fig:Ical}(a)) it is evident that current measurement with sub--\SI{5}{\nano\ampere} precision is possible in this system, which is much better than the design requirements.

\section{Testing with SiPMs}
The power supply has been independently tested so far. However, it is intended for use within a SiPM setup. And therefore, testing with SiPMs is necessary to evaluate the device. Consequently, the performance of this board is evaluated against a Keithley 2450 SMU using LED sources and dark noise using a Hamamatsu S13360-2050VE SiPM.

\subsection{Testing with LED pulses\label{sec:led_test}}
\begin{figure}[htbp]
    \centering
    \includegraphics[width=\textwidth]{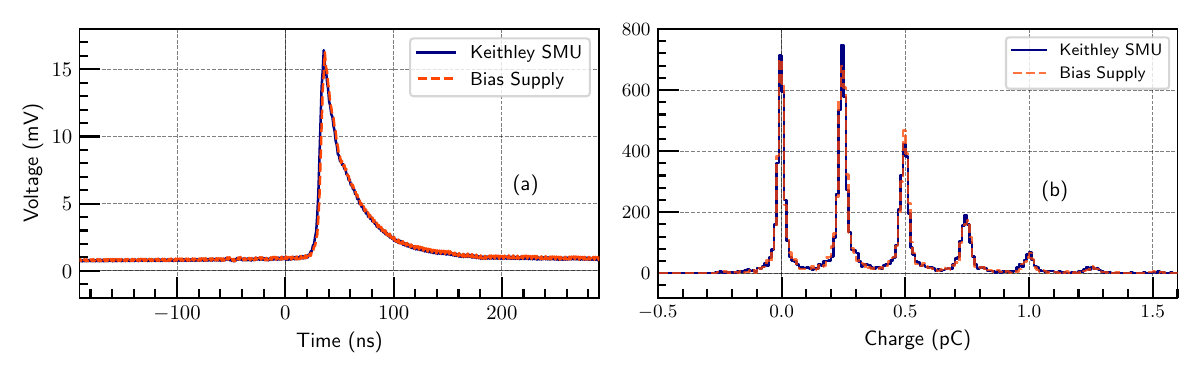}
    \caption{(a) Average pulse shape from Keithley SMU and the custom bias supply before baseline correction. (b) LED spectra of the SiPM. In both the plots, data for both the devices are superimposed on each other for a comparison.}
    \label{fig:led}
\end{figure}

We tested the power supply circuit with a CAEN SP5601 LED source \cite{caen_sp5601}. The signals were recorded using an oscilloscope and was triggered using the internal periodic trigger of the LED source. The LED source, SiPM, and the amplifier board are kept in a dark box, and the intensity of LED source was selected to have $\sim$1\,photoelectron (p.e.) per pulse. A total of 10,000 events were recorded for both bias sources, and the average pulse shape from the SiPMs are shown in figure \ref{fig:led}(a). The SiPM charge is estimated by integrating the voltage signal in a time window of \SIrange{20}{120}{\nano\second}, denoted as $t_i$ and $t_f$ in equation \ref{eqn:int_charge}, respectively, and dividing by the equivalent resistance of the transimpedance amplifier ($R$), and is represented as:

\begin{equation}
    q(t) = \frac{1}{R} \int_{t_i}^{t_f} V(t) dt
    \label{eqn:int_charge}
\end{equation}

The details of the setup and analysis method are described in \cite{Mamta_1}. Here, R\,=\,\SI{909.09}{\ohm} and $dt$ is \SI{1}{\nano\second}. The integration range covers more than 95\% of the signal. To avoid any offset due to the baseline change in events, the voltages in the time window -200 to \SI{-100}{\nano\second} are also integrated using equation \ref{eqn:int_charge} and are subtracted from the integrated signal. The integrated charge spectra of the same setup with two different bias supplies are shown in figure \ref{fig:led}(b). These two spectra look identical within the statistical fluctuations for all the properties of the SiPM signals, e.g.,

\begin{itemize}
    \item Pedestal position and width,
    \item The shape and width of pulses for different number of photo-electrons,
    \item Relative number of p.e. in the spectrum, and
    \item The baseline of the continuous background levels due to correlated noises.
\end{itemize}

In summary, the properties of the LED signal using the custom board are similar to those of the commercial power supply (Keithley 2450 SMU).

\subsection{Noise rate measurement of SiPMs}
\begin{figure}[htbp]
    \centering
    \includegraphics[width=\textwidth]{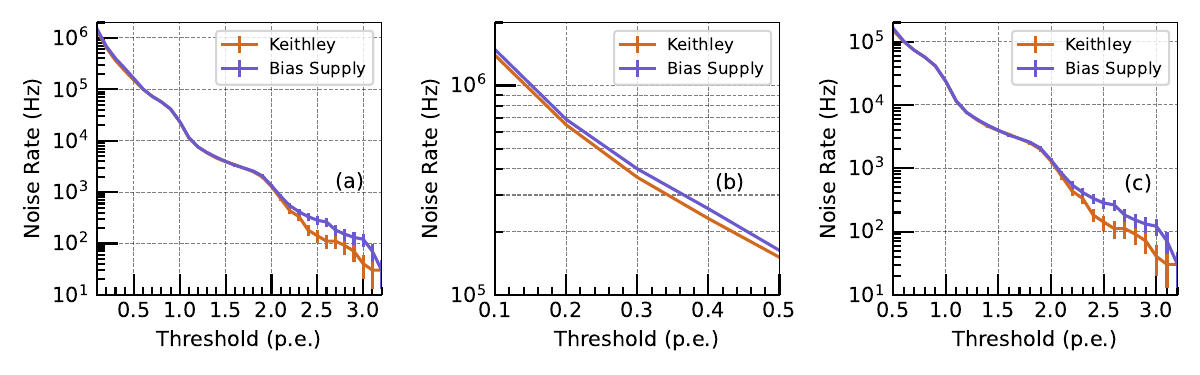}
    \caption{Comparison of noise rates of the SiPM: (a) entire range (0.1 p.e. to 3.2 p.e.), (b) \textless 0.5 p.e., and (c) \textgreater 0.5 p.e.}
    \label{fig:noise}
\end{figure}

The bias supply has been compared with a Keithley 2450 SMU for dark counts of the SiPM, similar to what was done by Jangra et al. \cite{Mamta_1}. The signals were recorded using an oscilloscope with a periodic trigger from a signal source. The samples were taken every \SI{1}{\nano\second} for a period of \SI{10}{\micro\second}, that is, 10,000 sample points per event. A total of 10,000 such events were recorded. Each of the events was divided into 100 sections corresponding to \SI{100}{\nano\second} intervals and numerically integrated using equation \ref{eqn:int_charge} to obtain the charge. The first \SI{100}{\nano\second} section of all events was considered as the baseline and was subtracted from the other 99 sections for the baseline correction. These charge values were histogrammed for the calculation of the noise rates. The cumulative distribution of the counts at different charge values gives us the noise rate. These are shown in figure \ref{fig:noise}(a). The figures \ref{fig:noise}(b) and \ref{fig:noise}(c) show the same, but zoomed in two different ranges: \textless 0.5 p.e. and \textgreater 0.5 p.e., respectively. The counts corresponding to \textless 0.5 p.e. primarily denote the electronic noise. The value of the charge depends on the gain of the device, as discussed in Section \ref{sec:theory}, and the charge corresponding to 1 p.e. can be found from the gain (calculated from figure \ref{fig:led}(b)).

The DCR is defined as the noise rate above charge corresponding to 0.5\,p.e. The noise rates are given in table \ref{tab:noise_rates}. For CMVD the valid event selection criteria is set to \textgreater 2.5\,p. e. From the values of the dark count rates, it is evident that both devices perform similarly and are consistent with each other within the margin of error, under consideration of the CMVD event selection criteria. Due to low statistics, the noise rate above 2.5\,p.e. deviates $\sim2\,\sigma$.

\begin{table}[h]
    \centering
    \begin{tabular}{cccc}
        \toprule
        \multirow{2}{*}{Device} & \multicolumn{3}{c}{Noise Rate (kHz)}                                                   \\
                                & \textgreater 0.5\,p.e.               & \textgreater 1.5\,p.e. & \textgreater 2.5\,p.e. \\
        \midrule
        Keithley 2450 SMU       & $151.101 \pm 1.226$                  & $3.909 \pm 0.199$      & $0.141 \pm 0.038$      \\
        Custom bias supply      & $162.990 \pm 1.273$                  & $3.970 \pm 0.200$      & $0.283 \pm 0.053$      \\
        \bottomrule
    \end{tabular}
    \caption{Comparison of noise rates.}
    \label{tab:noise_rates}
\end{table}

\section{Justification of the electrical specifications}
To ensure reliable detection of the scintillation light, specifically for low photon counts expected from cosmic muons, the gain of the SiPMs must remain sufficiently stable, as discussed previously. Since the SiPM gain ($G$) depends linearly on the applied overvoltage (as shown in equation \ref{eqn:gain}), fluctuations in the bias voltage impact the gain equally by fraction--

\begin{align}
    G                     & = G_0 \cdot V_{OV} \nonumber \\
    \implies \frac{dG}{G} & = \frac{dV_{OV}}{V_{OV}}
    \label{eqn:dGbyG}
\end{align}

The SiPMs are operated at $V_{OV}\sim$\SI{3}{\volt}. Therefore, according to equation \ref{eqn:dGbyG}, a fluctuation in the overvoltage of 2 ADC counts (as shown in figure \ref{fig:PIDcal}(c)-(g) and figure \ref{fig:Vcal}(b)), that is, $dV_{OV}=$\SI{6}{\milli\volt}, will result in a gain fluctuation of--
\begin{align}
    \frac{dG}{G} & = \frac{dV_{OV}}{V_{OV}} \nonumber                          \\
                 & = \frac{\SI{6}{\milli\volt}}{\SI{3}{\volt}} = 0.002 = 0.2\%
\end{align}

However, from the tests in section \ref{sec:led_test} and the charge spectra of figure \ref{fig:led}(b), the gain fluctuation is approximately 5\% when taking care of the peak widths. This is much larger than the 0.2\% expected from the bias supply circuit. Therefore, stability requirement of the bias supply for the CMVD is met.

While finer control is technically possible, choice of \SI{50}{\milli\volt} for the step size represents a practical compromise between electronics, firmware and calibration complexity, and control precision. More importantly, all 32 SiPMs within a group are supplied from a common source and have been grouped based on closely matched breakdown voltages (as discussed in section \ref{sec:requirements}). This ensures that all devices receive nearly the same overvoltage, thereby maintaining gain uniformity across the group. In practice, the relative difference in overvoltage across grouped SiPMs is well below the \SI{50}{\milli\volt} step and thus does not introduce significant gain mismatch.

Based on device characteristics, a single SiPM draws $\sim$\SI{50}{\nano\ampere} at $V_{OV}\sim$\SI{3}{\volt}. Consequently, a resolution of \SI{50}{\nano\ampere} was set as the maximum least count allowed for current measurement, in order to detect the presence of individual SiPMs. However, figure \ref{fig:Ical}(a) shows that the achieved current resolution was better than \SI{5}{\nano\ampere}, which is an improvement of an order of magnitude, allowing precise monitoring of the current and offering diagnostic capabilities in the case of problems, malfunctions, or failures.

\section{Discussion and conclusions}
Although several similar devices are available on the market, this design provides us with the flexibility to customize the voltage range, step size, and noise performance according to our requirements. The design is modular, allowing for easy integration into larger systems like the CMVD. The PID control algorithm implemented in the MCU provides a stable output voltage with low noise, as demonstrated by the tests. A comprehensive review of the different devices on the market and a custom-made one is given in table \ref{tab:comparison}. It is also worth mentioning that the custom design is significantly cheaper than the commercially available devices, making it a cost-effective solution for our application.

\begin{table}[htbp]
    \centering
    \caption{Comparison of SiPM bias supplies}
    \begin{tabular}{|l|l|c|c|c|c|c|}
        \hline
        \multicolumn{2}{|l|}{\textbf{Device Parameter}}                        &
        \makecell{\textbf{Hamamatsu}                                                                   \\ \textbf{C11204-01}\\ \cite{hamamatsu_bias}} &
        \makecell{\textbf{CAEN}                                                                        \\ \textbf{A7585D/DU}\\ \cite{caen_bias}} &
        \makecell{\textbf{AiT}                                                                         \\ \textbf{HV80B}\\ \cite{AiT_bias}} &
        \makecell{\textbf{Gil}                                                                         \\ \textbf{et al.}\\ \cite{gil_programmable_2011}} &
        \makecell{\textbf{This}                                                                        \\ \textbf{work}} \\
        \hline

        \multicolumn{2}{|l|}{\makecell[l]{\textbf{Voltage}                                             \\ \textbf{range}}} &
        \SIrange[range-phrase = --, range-units = single]{20}{90}{\volt}       &
        \SIrange[range-phrase = --, range-units = single]{20}{85}{\volt}       &
        \SIrange[range-phrase = --, range-units = single]{10}{80}{\volt}       &
        \SIrange[range-phrase = --, range-units = single]{0}{200}{\volt}       &
        \SIrange[range-phrase = --, range-units = single]{50}{58}{\volt}                               \\
        \hline
        \multicolumn{2}{|l|}{\textbf{Ripple}}                                  &
        \SI{0.1}{\milli\volt}\textsubscript{p-p}                               &
        \SI{0.2}{\milli\volt}\textsubscript{p-p}                               &
        \SI{0.5}{\milli\volt}\textsubscript{RMS}                               &
        \SI{40}{\milli\volt}\textsubscript{p-p}                                &
        \SI{10}{\milli\volt}                                                                           \\
        \hline

        \multirow{2}{*}{\makecell[l]{\textbf{Voltage}                                                  \\ \textbf{setting}}} & \textbf{Method} &
        Digital                                                                &
        Digital                                                                &
        Analog                                                                 &
        Digital                                                                &
        Digital                                                                                        \\
        \cline{2-7}
                                                                               & \textbf{Resolution} &
        \SI{1.8}{\milli\volt}                                                  &
        \SI{1}{\milli\volt}                                                    &
        NA                                                                     &
        \SI{2.5}{\milli\volt}                                                  &
        \makecell[c]{\SI{50}{\milli\volt}                                                              \\ (can go\\ lower)} \\
        \hline

        \multicolumn{2}{|l|}{\makecell[l]{\textbf{Max. output}                                         \\ \textbf{current}}} &
        \SI{2}{\milli\ampere}                                                  &
        \SI{10}{\milli\ampere}                                                 &
        \SI{4}{\milli\ampere}                                                  &
        \SI{5}{\milli\ampere}                                                  &
        \SI{1}{\milli\ampere}                                                                          \\
        \hline

        \multirow{3}{*}{\makecell[l]{\textbf{Current}                                                  \\ \textbf{monitor}}} & \makecell[l]{\textbf{Readout} \\ \textbf{method}} &
        None                                                                   &
        \makecell[c]{Analog +                                                                          \\ Digital} &
        analog                                                                 &
        None                                                                   &
        Digital                                                                                        \\
        \cline{2-7}
                                                                               & \textbf{Range}      &
        NA                                                                     &
        full range                                                             &
        \SIrange[range-phrase = --, range-units = single]{0}{5}{\milli\ampere} &
        NA                                                                     &
        \SIrange[range-phrase = --, range-units = single]{0}{5}{\micro\ampere}                         \\
        \cline{2-7}
                                                                               & \textbf{Resolution} &
        NA                                                                     &
        \makecell[c]{high range:                                                                       \\ \quad\SI{650}{\nano\ampere} \\ low range:\\ \quad\SI{100}{\nano\ampere}} &
        NA                                                                     &
        NA                                                                     &
        \textless \SI{5}{\nano\ampere}                                                                 \\
        \hline
    \end{tabular}
    \label{tab:comparison}
\end{table}

From the comprehensive tests of the custom bias supply, we can conclude that this design is suitable for the CMVD at TIFR, Mumbai. The stability and noise performance are as per the requirements of the experiment. In principle, the stability may be further improved by better tuning the PID controller.

The CMVD is being built in an air-conditioned laboratory where the temperature is regulated within \textpm\SI{1}{\degreeCelsius} and there are no large--scale temperature fluctuations, and hence the temperature compensation of the bias voltage is not critical. However, due to the modular nature of the device, temperature control can be incorporated without any difficulties, as discussed in Section \ref{sec:design}.

The primary drawback of this circuit is that it has been designed for a voltage range of \SIrange[range-units = single]{50}{58}{\volt}, which is specific to the particular make and model of SiPMs that will be used in the CMVD. We have already begun developing an improved design that can provide a much broader range of voltages, along with enhanced stability and noise performance.

\acknowledgments
We want to thank Darshana Gonji for assembling the PCBs. The authors thank Raj Shah and Dr. Mamta Jangra for their help in testing the device. The authors also thank Abhijeet Ghodgaonkar for his input regarding the circuit design.

\bibliographystyle{JHEP}
\bibliography{biblio}

\end{document}